\let\a=\alpha \let\b=\beta  
\let\d=\delta  \let\e=\varepsilon
 \let\g=\gamma \let\h=\eta \let\k=\kappa  \let\l=\lambda
      \let\o=\omega
  
  \let\s=\sigma 

  \let\z=\zeta

\let\D=\Delta

\let\i=\infty

\def\oo{{\o}}  
\def\xx{{\vec x}} \def\yy{{\vec y}} \def\kk{{\vec k}}

\def\LL{{\cal L}}

\def\DD{{\cal D}}\def\GG{{\cal G}}

\def\nn{\nonumber}

\documentstyle[twocolumn]{article}
\begin{document}


\title{Anomalous BCS equation 
        for a Luttinger superconductor}
\author{V.Mastropietro\\
        Dipartimento di Matematica, Universit\`a ``Tor Vergata'' di Roma}
\maketitle
\begin{abstract}
{\it In the context of the Anderson theory of high $T_c$
cuprates, we develop
a BCS theory for Luttinger liquids.
If the Luttinger interaction is much
stronger than the BCS potential we find
that the BCS equation
is quite modified compared to usual 
BCS equation for Fermi liquids.
In particular $T_c$
predicted by the BCS equation for Luttinger liquids
is quite higher than the usual $T_c$
for Fermi liquids.}
\end{abstract}
\vskip.5cm
\section{The anomalous BCS equation}
The equivalent of BCS theory for a Luttinger liquid
has not formally worked out, despite the relevance
of such theory for the problem of superconductivity
in the high-$T_c$ cuprates, see [1] 
or the discussion in the following section. 
Such theory should describe superconductors which in their normal state
are Luttinger liquids.
In this paper we
develop such theory using constructive
quantum field theory
techniques, applied in many other Luttinger liquid problems,
see [2], [3]. In particular, we compute in a
rigorous way the BCS self-consistence equation for a spinning
Luttinger liquid (the {\it Mattis model}, see [4]) 
coupled with a BCS anomalous potential.
We stress that 
it should be difficult to find similar results
using different techniques: in fact the Mattis model plus
a BCS interaction is not exactly solvable, and bosonization or conformal
quantum field theory cannot be used
to compute the correlation functions of theories with gap like the present one
(see for instance
[5]).

We will assume\footnote{See [1] pag.209}
that an anomalous
self energy has been introduced by some external influence
in a Luttinger liquid so that
the model is described by the following hamiltonian
\begin{equation}\label{1}
H=H_a+H_{BCS}
\end{equation}
where $H_a$ is the {\it Mattis model} hamiltonian 
\begin{eqnarray}\label{2}
&&{2\pi\over L}\sum_{k,\o,\s} (\o k-p_F)\psi^{+}_{k,\o,\s}
\psi^{-}_{k,\o,\s}-\l [{2\pi\over L}]^3\sum_{k_1,k_2,p}\\
&&\sum_{\o,\s,\s'}v(p)
:\psi^{+}_{k_1,\o,\s}\psi^{-}_{k_1-p,\o,\s}:
:\psi^{+}_{k_2,-\o,\s'}\psi^{-}_{k_2+p,-\o,\s'}:\nn
\end{eqnarray}
where $\psi^{\pm}_{k,\o,\s}$ are fermionic creation or
annihilation operators with momentum\footnote{We are considering
the Mattis model on a lattice, while the original
model introduced and solved in [4] is defined on the continuum} 
$k={2n\pi/L}$, $-L/2\le n\le
{L-1\over 2}$, quasi-particle index $\o=\pm 1$,
spin $\s=\pm {1\over 2}$; $|v(p)|\le e^{-\k|p|}$. {\it We assume that $\l>0$}.
The BCS interaction is described as usual, {\it if $g>0$}
\begin{eqnarray}
&&H_{BCS}=-\D {2\pi\over L}\sum_{k,\o}
\psi^+_{k,\o,{1\over 2}}
\psi^+_{-k,-\o,-{1\over 2}}\nn\\
&&-\D {2\pi\over L}\sum_{k,\o}
\psi^-_{-k,-\o,-{1\over 2}}
\psi^-_{k,\o,{1\over 2}}+{\D^2\over g}
\end{eqnarray}
We assume that $\D\in R$ and $\D\ge 0$. Note that
if $\D=0$ the model is exactly solvable [4]; the system is a Luttinger
liquid (it is perhaps the simplest spinning model showing
Luttinger liquid behaviour) and shows {\it spin-charge separation}.

The ground state energy $E_0(\D)$ depends on $\D$; the BCS equation is
the extremizing equation ${\partial E_0(\D)\over \partial\D}=0$ 
which
has the form (see [6] for the proof of a similar statement):
\begin{equation}\label{BCS}
\D=g{2\pi\over L}\sum_{k,\o} <\psi^+_{k,\o,{1\over 2}}
\psi^+_{-k,-\o,-{1\over 2}}>
\end{equation}
If $\l=0$ the r.h.s of eq.(\ref{BCS}) can be calculated very easily,
as the hamiltonian can be put in diagonal form by performing a {\it Bogolubov
transformation}. In the interacting case an exact solution is not
available and we compute the r.h.s. of eq.(4) writing it as a
convergent series
for $\D,\l$ suitably small {\it i.e.} $\D,\l<\e$,
if $\e$ is a suitable number $<<1$
(but no restriction is imposed
on their ratio). Note that the restriction on the smallness
of $\D,\l$ is done only for technical reasons
in order to ensure convergence of the perturbative series.
However it is very likely that our 
bound for the convergence radius are far from to be optimal,
and instead our results hold also for $\l=O(1)$; if $\D=0$
we know that this is true directly from the exact solution [4]
\footnote{The role of the lattice should play no role.}.

Our main result is
\vskip.5cm
{\it There exists an $\e>0$ such that, for $\D, \l\le\e$ 
the BCS equation 
(\ref{BCS}), 
in the limit $\b\to\i, L\to\i$, can be written as
\begin{equation}\label{sper}
{a\D\over g}={\D\over \h} [|\D|^{-\h}-A^{-\h}](1+\l f)
\end{equation}
where $\h=\b_1\l+O(\l^2)$ is a critical index,
$f\equiv f(\l, \D)$, $|f|\le {\rm const}$, and $A$,
$a,\b_1$ are positive constants.}
\vskip.5cm

Eq.(\ref{sper}) is difficult to solve in general; we study its non trivial solutions
for two limiting cases, denoting 
by $\s(x)$ a function bounded by $|x|$ times a
constant.

{\bf 1)} ${\l\over g}<<1$, which means that the BCS hamiltonian is
{\it weakly} perturbed by the Luttinger interaction. As one expects, 
the gap
is given by
\begin{equation}
\D=A e^{-{a\over g}(1+\s(\l))}
\end{equation}
so that the gap is exponentially small and the Luttinger liquid 
behaviour of the
system plays no role. 

{\bf 2)} ${\l\over g}>>1$, which means that the Mattis hamiltonian 
is {\it
weakly}
perturbed by the BCS interaction.
We find, remembering that $\h=\b_1\l+O(\l^2)$
\begin{equation}
\D=A[{g\over a\h}]^{1\over \h}[1+\s(\l)+\s({g\over\l})]^
{1\over \h}
\end{equation}
We see that the
{\it Luttinger interaction enhances strongly
the gap} with respect to the $\l=0$ case
if ${\l\over g}>>1$ 
\begin{equation}\label{m}
\D\simeq B e^{-{a\over g}[{g\over a\h}|\log {g\over a\h}|]}>>
B' e^{-{a\over g}}
\end{equation}
{\it Note also the crucial role of the {\it sign} of the Luttinger
interaction}.

\section{A model for the high-$T_c$ cuprates}
In the previous section we have written a BCS equation for
a Luttinger liquid and we have seen that, if the Luttinger
interaction is much stronger than the BCS interaction,
there is a large deviation from 
the Fermi liquid BCS equation and in
particular the gap is 
much larger. Can this result be applied to the physics
of high-$T_c$ cuprates?
 
The Anderson theory of high-$T_c$ cuprates is based on the following
points\footnote{See for instance pag.46-55 of [1]}:

1) the conduction electrons are confined on layers;

2)the interaction between electrons in the same layer is much
stronger
than the interlayer interaction;

3) the normal state of the electrons on the layers 
is a Luttinger liquid;

4) the interlayer
pairing allows Cooper pairs to tunneling into an adiacent layer
by the Josephson mechanism\footnote{See also pag. 308 of [1]}

The interlayer interaction is assumed to be given
by the {\it Josephson pair tunneling hamiltonian}\footnote{See 
for instance eq.(7)
pag 321 of [1], or eq.(9) (in $d=2$) below} 
and some physical arguments
for
motivating this choice
and for not considering single particle tunneling are
given\footnote{See for instance the considerations 
based on the holon-spinon coherence
at pag.50 of [1] or in
the Clarke reprint [o]}. The model is then studied
by the usual {\it mean field BCS} approximation\footnote{see eq.(16)
pag.218 of [1] or the considerations after eq.(10) below}.
However by point 3)
one has to take into account in
the resulting  BCS equation the Luttinger liquid nature
of the fermions in their normal state. 
This is a big problem as no theory showing
the Luttinger liquid behaviour of bidimensional strongly interacting
fermions exists\footnote{In [1] this is essentially derived
from experimental results} so that
the form of
a Luttinger liquid
BCS equation is essentially {\it guessed} by replacing
in the usual BCS the Fermi propagator with the one dimensional
Luttinger model propagator\footnote{See pag.213 eq.(12) of [1]}.

As the above theory is a {\it mean field} theory, its results
should be, as usual, {\it independent} from the dimensions;
this means that we can expect that coupling
$d=1$ or $d=2$ Luttinger liquids and performing a BCS approximation
on the pair tunnelling interaction, the results are
qualitatively
the
same.
In other words {\it we expect that the predictions
of a BCS theory
for a superconductor whose normal state is a Luttinger liquid
are qualitatively the same whatever its dimensions are}. 
This is just
what happens for the usual superconductors
whose normal
state is a Fermi liquid; the BCS approximation leads
for instance
to a gap or a critical temperature
exponentially small in $g$ in {\it any
dimension}.
The crucial and very non trivial assumption
is that there exist $d=2$ fermions with a Luttinger liquid
behaviour in their normal state; but, {\it if} they exists,
the results of a BCS theory of coupled
Luttinger liquids should be essentially indipendent on
their dimensions. 

Considering one dimensional coupled Luttinger liquids, 
the BCS equation can computed in a rigorous way.
The Luttinger
liquids can be described
by the Mattis model hamiltonian eq.(2), giving an extra {\it chain index} to the Fermi 
operators $i=a,b$. The two Mattis hamiltonians will be called
$H_a$ and $H_b$.
The {\it pair-hopping} hamiltonian
$H_{int}$ is, following [1]\footnote{See eq.(7) pag.320}
\begin{eqnarray}\label{3}
&&H_{int}=-g [{2\pi\over L}\sum_{k,\o} \psi^+_{k,\o,{1\over 2},a}
\psi^+_{-k,-\o,-{1\over 2},a}]\\
&&[{2\pi\over L}\sum_{k',\o} \psi^-_{-k',-\o,-{1\over 2},b}
\psi^-_{k',\o,{1\over 2},b}]+{\rm c.c.}\nn
\end{eqnarray}
The total hamiltonian is $H_a+H_b+H_{int}$.
Let we write in eq.(\ref{3})
$${2\pi\over L}\sum_{k,\o} \psi^+_{k,\o,{1\over 2},i}
\psi^+_{-k,-\o,-{1\over 2},i}$$ as 
\begin{eqnarray}
&&{2\pi\over L}\sum_{k,\o} <\psi^+_{k,\o,{1\over 2},i}
\psi^+_{-k,-\o,-{1\over 2},i}>+\label{f}\\
&&{2\pi\over L}\sum_{k,\o}[
\psi^+_{k,\o,{1\over 2},i}
\psi^+_{-k,-\o,-{1\over 2},i}-<\psi^+_{k,\o,{1\over 2},i}
\psi^+_{-k,-\o,-{1\over 2},i}>]\nn
\end{eqnarray}
The {\it BCS approximation} consists in
replacing $H_{int}$ with $H_{int}^{app}$ obtained
neglecting
the terms bilinear in the "fluctuations" {\it i.e.} in
the second addend
in eq.(\ref{f}).
We obtain\footnote{See eq.(16) pag.218 of [1]}
$$H_{int}^{app}=-\D {2\pi\over L}\sum_{k,\o}
\psi^+_{k,\o,{1\over 2},a}
\psi^+_{-k,-\o,-{1\over 2},a}$$
$$-\D {2\pi\over L}\sum_{k,\o}
\psi^-_{-k,-\o,{-1\over 2},b}
\psi^-_{k,\o,{1\over 2},b}+{\D^2\over g}+{\rm c.c.}$$
and self-consistency requires
eq.(\ref{BCS}). 
Replacing $H_{int}$ with $H_{int}^{app}$\footnote{
The above computation follows the Anderson gap derivation, 
see chap. 7 in [1],
but for semplicity
we have neglected the electron phonon
interaction, see eq.(21) pag. 219 of [1]} 
has the effect that
the model is described
by two indipendent hamiltonians, $H=\tilde H_a+\tilde H_b$,
each one given by eq.(1).

The BCS equation for coupled Luttinger liquids is then given by
eq.(5). In the range of parameters physically reasonable (see point 2)
above) {\it i.e.} if $\l>>g$ and noting that, as usual, $T_c$
is proportional to $\D$ (see below) we find that $T_c\simeq \tilde B 
e^{-{a\over g}[{g\over a\h}
|\log{g\over a\h}|]}$ {\it i.e.} {\it much higher} than
$T_c$ for normal superconductor $T_c\simeq \tilde B'e^{-{a\over g}}$. 

In order to compare our result with [1],
note that the r.h.s. of eq.(4) computed\footnote{see eq.(12) pag.213} 
in [1] is similar to our 
eq.(\ref{ma}) below (see especially the presence of the wave
function renormalization) but $\s_h\simeq \D \g^{-(\b_1\l+..)h}$ 
is replaced by $\D$ {\it i.e.}
the {\it anomalous flow of the BCS gap is neglected}. Such
effect is on the other hand {\it crucial} for our analysis: 
all the conclusion drawn from eq.(5) in the preceding section
are based on the fact that $\b_1\not=0$. In fact it is this anomalous
enlarging of the gap due to the Luttinger interaction
which produces a much larger solution of the BCS equation.
On the other hand 
the gap renormalization in Luttinger liquids is a 
well known fact, see for instance the 
gap of the $XYZ$ model, [9].

Finally we stress that one can try to study directly
a model of coupled chains, see [7], 
without any approximation, and
it could happen that the model
is not really described by our BCS approximation.
This has not real relevance for our analysis, as our aim
was just to find a BCS equation for coupled system of fermions  
on planes with a Luttinger liquid behaviour,
and we make a BCS computation in $d=1$ using the fact that a mean field theory
like BCS should be {\it insensitive} to the dimensions.
So it could be that our BCS equation eq.(5) 
could be applied to coupled planes and not chains,
and on the other hand a detailed analysis of the model given of eq.(10)
could give no insight on the problem of coupled planes as the behavior
of the system depend on the dimensions.
This is in fact just what happens
in the usual 
BCS theory: the BCS equation is qualitatively the same
in any dimension, but in $d=1$ the mean field approximation 
is not correct. The real question is if really bidimensional
Luttinger exists, but this question is not addressed here and we refer to [1].

\section{Renormalization group analysis}
We discuss now how to compute the r.h.s. of eq.(\ref{BCS}). 
We introduce as usual a set of {\it Grassman variables} 
$\psi^{\pm}_{\kk,\oo,\s}$, $\kk=(k_0,k)$
and $k={2\pi n_1\over L}$, $k_0={2\pi(n_0+2^{-1})\over\b}$,
if $n_0, n_1$
are integers and $-{L\over 2}\le n_1\le (L-1)/2$,
$-{\b\over 2}\le n_2\le {\b-1\over 2}$, if $1/\b$ is the temperature.
The {\it Grassmanian
integration} $P(d\psi)$ is defined by the anticommutative
Wick rule with propagator
$$g_\o(\kk)={1\over -ik_0+\o k-p_F}$$
In general we denote by $\int \{\DD \psi e^{-\int d\kk \psi^+_\kk
h(\kk)^{-1} \psi^-_\kk} \}$ the Grassmanian integration with propagator
$h(\kk)$, 
where $\int d\kk={(2\pi)^2\over
L\b}\sum_\kk$; in particular
$$P(d\psi)=\{\DD\psi \prod_{\kk,\o,\s}e^{ -\int d\kk
\psi^+_{\kk,\o,\s}(-ik_0+\o k-p_F)\psi^-_{\kk,\o,\s}}\}$$
Then we write the r.h.s. side of eq.(\ref{BCS}) as a functional integral
\begin{eqnarray}
&&<\psi^{+}_{\kk,1,{1\over 2}}\psi^{+}_{-\kk,-1,-{1\over 2}}>=\label{3a}\\
&&{1\over \int P(d\psi)
e^{-V(\psi)}}
\int P(d\psi)
e^{-V(\psi)}
\psi^{+}_{\kk,1,{1\over 2}}\psi^{+}_{-\kk,-1,-{1\over 2}}
\nn
\end{eqnarray}
where $V(\psi)=\l \bar V+\D P$
and 
$$P=\sum_\o\int d\kk (\psi^+_{\kk,\o,{1\over 2}}
\psi^+_{-\kk,-\o,-{1\over 2}}+
\psi^-_{-\kk,-\o,{-1\over 2}}
\psi^-_{\kk,\o,{1\over 2}})$$
\begin{eqnarray}
&&V=\int \prod_{i=1}^4 d\kk_i\d(\sum_{i=1}^4\e_i\kk_i)\nn\\
&&\sum_{\o,\s,\s'}:\psi^{+}_{\kk_1,\o,\s}
\psi^{-}_{\kk_2,\o,\s}:
:\psi^{+}_{\kk_3,-\o,\s'}\psi^{-}_{\kk_4,-\o,\s'}:\nn
\end{eqnarray}
We evaluate the above Grassman integral using Wilsonian renormalization group
techniques. It is convenient to write $k=k'+\o p_F$,
where $k'$ is the momentum measured from the Fermi surface.

We decompose the integration $P(d\psi)$ into a product
of independent integrations. 
This can be done writing
\begin{equation}\label{dec}
g_\o(\kk'+\o
p_F)=\sum_{h=0}^{-\i}g_\o^{h}(\kk'+\o
p_F)+g^1_\o(\kk'+\o p_F)
\end{equation}
with 
$$g^h_\o(\kk'+\o p_F)={f_h(\kk')\over -ik_0+\o k'}$$
and $f_1(\kk')=1-\chi(\kk')$, 
$\chi(\kk')\equiv\bar \chi(|\kk'|)$ 
is a smooth compact support function
such that $\chi(\kk')=1$ for $|\kk'|\le \g^{-1}$ and $\chi(\kk')=0$ for
$|\kk'|\ge 1$, if $\g>1$; moreover for $h\le 0$
$$f_h(\kk')=\chi(\g^{-h}\kk')-\chi(\g^{-h+1}\kk')$$
is a smooth compact support function 
non vanishing only for $\g^{h-2}\le |\kk'|\le \g^{h}$. 
Then $g^1_\o(\kk'+\o p_F)$ is the {\it ultraviolet} part of the propagator,
while $\sum_{k=0}^{-\i} g^h_\o(\kk')$ is the infrared part. Note that,
from the compact support properties of $g_\o^h$, the sum
in eq.(12) is from $0$ to $h_\b$, where $\g^{h_\b}={\pi\over\b}$,
as $|k_0|\ge {\pi\over\b}$. Let be $C_h^{-1}=\sum_{k=h_\b}^h f_k$.

The ultraviolet integration is somehow special (and essentially trivial
for the presence of the lattice)
and we will not discuss it
here, see [2].
If $\l=0$ the infrared integration can be done by
performing the well known
{\it Bogolubov transformation} to diagonalize the BCS
hamiltonian. If $\l\not=0$
the BCS gap and
the wave function renormalization 
have a non trivial RG flow so that we have to perform
a {\it different} Bogolubov transformation at each iteration of the RG.
We set
$Z_0=1$: once the fields $\psi^{(0)},\ldots,\psi^{(h+1)}$ have been
integrated we have:
\begin{eqnarray}\label{11}
&&\int\{ {\cal D}\psi^{(\leq h)}
\prod_{\o=\pm 1}e^{-\int d\kk' C_h Z_h \vec\psi^{(\leq
h)+}_{\kk',\o}\GG_\oo^{(h)}(\kk')^{-1}
\vec\psi^{(\leq h)-}_{\kk',\o}}\}\nn\\
&&e^{-V^h(\sqrt{Z_h}\psi^{(\leq h)})}\label{b}
\end{eqnarray}
if
$$\vec\psi_{\kk',\o}^{(\le h)\pm}=(\psi^{(\le h)\pm}_{\kk'+\o p_F,\o,{1\over2}},
\psi^{(\le h)\mp}_{-\kk'-\o p_F,-\o,-{1\over 2}})$$
and $\GG^{(h)}(\kk')^{-1}$ is defined by

\begin{displaymath}\label{7a}
\left(
\begin{array}{c c}
-ik_0+\o k'& \s_{h}(k')\\
\s_h(k')&-ik_0-\o k'
\end{array}\right)
\end{displaymath}

$V^h$ is called the {\it effective potential} at scale h and is given by
\begin{eqnarray}
&&V^h(\psi^{(\le h)})=\sum_{n=2}^\i\int \prod_{i=1}^{n} d\kk_i\\
&&W_{n}^h(\kk_1,\ldots,\kk_{n})\prod_{i=1}^{n}\psi^{(\le h)\e_i}_{\kk_i,\oo_i,\s_i}
\d(\sum_{i=1}^n\e_i (\kk'_i+\oo _i p_F))\label{cc}\nn
\end{eqnarray}

We define a {\it localization operator} $\LL$
extracting the relevant or marginal
part
of the {\it effective potential} $V^h$:

i) If $n>4$ then $\LL W^h_n=0$;

ii) Let be $n=4$. In this case $\LL W^h_4=0$
unless $\sum_{i=1}^4\e_i\o_i
p_F=0$, $\sum_{i=1}^4\e_i=0$
in which case the action is non trivial and it is given by
\begin{equation}\label{loc1}
\LL W^h_4(\kk'_1+\oo_1 p_F,....)=
W_{4}^{h}(\o_1 p_F,...)
\end{equation}

iii) if $n=2$ then if $\sum_{i}\e_i=0$
\begin{eqnarray}\label{loc2}
&&\LL W_{2}^{h}(\kk'_1+\o_1 p_F,
\kk'_2+\o_2 p_F)=[W_{2}^{h}(\o_1 p_F,\o_2 p_F)+\nn\\
&&\o_1E(k'+\o_1 p_F)
\partial_{k}
W_{2}^{h}(\o_1 p_F,\o_2 p_F)
+k^0\nn\\
&&\partial_{k_0}W_{2}^{h}(\o_1 p_F,\o_2 p_F)]
\end{eqnarray}

while if $\sum_i\e_i\not=0$
then
$$\LL W_{2}^{h}(\kk'_1+\o_1 p_F,
\kk'_2+\o_2 p_F)=W_{2}^{h}(\o_1 p_F,\o_2 p_F)$$

We can write then the relevant part of the effective potential as:
\begin{equation}
\LL V^{h}=\g^h n_h F_\nu^{h}+s_h F_\s^{h}+z_h F_\z^{
h}+a_h F_\a^{h}+g_{2,h} F_{2}^{h}+g_{4,h} F_{4}^{h}
\end{equation}
where
\begin{eqnarray}
&&F_i^{h}=\sum_\o \int d\kk'
 f_i \psi^{(\le h)+}_{\kk'+\o p_F,\o,\s}
\psi^{(\le h)-}_{\kk'+\o p_F,\o,\s}\nn\\
&&F_\s^{h}=\sum_\o \int d\kk'
\psi^{(\le h)+}_{\kk'+\o p_F,\o,{1\over 2}}
\psi^{(\le h)+}_{-\kk'-\o p_F,-\o,-{1\over 2}}+{\rm c.c.}\nn\\
&&F_{2}^{h}=\int [\prod_{i=1}^4 d\kk_i\d(\sum_i \kk_i)\sum_{\o,\s,\s'}\nn\\
&&[\psi^{(\le h),+}_{\kk'_1+ \oo p_F,\oo,\s}
\psi^{(\le h),-}_{\kk'_2+\oo p_F,\oo,\s}]
[\psi^{(\le h),+}_{\kk'_3-\oo p_F,-\oo,\s'}
\psi^{(\le h),-}_{\kk'_4-\oo p_F,-\oo,\s'}]\nn\\
&&F_{4}^{h}=\int [\prod_{i=1}^4 d\kk_i\d(\sum_i \kk_i)\sum_{\o,\s,\s'}\nn\\
&&[\psi^{(\le h),+}_{\kk'_1+ \oo p_F,\oo,\s}
\psi^{(\le h),-}_{\kk'_2+\oo p_F,\oo,\s}]
[\psi^{(\le h),+}_{\kk'_3+\oo p_F,\oo,\s'}
\psi^{(\le h),-}_{\kk'_4+\oo p_F,\oo,\s'}]\nn
\end{eqnarray}

where $i=\nu,\a,\z$, $f_\nu=1$, $f_\a=\o k'$,
$f_\z=-i k_0$. Moreover
$g_{2,0}=\hat v(0)\l+O(\l^2)$, $|g_{4,0}|\le C\l^2$,
$s_0=\D+O(\D \l)$, $a_0,z_0=O(\l)$,
$n_0=O(\D\l)$.
We write eq.(\ref{11}) as:
\begin{eqnarray}\label{neo6}
&&\int {\cal D}\psi^{(\leq h)}
e^{-\int dk' \vec\psi^{(\leq
h)+}_{k'}C_h Z_{h-1}(k')\GG^{(h-1)}(k')^{-1}
\vec\psi^{(\leq h)-}_{k'}}\nn\\
&&e^{-\tilde V^h(\sqrt{Z_h} \psi^{(\leq h)})}]
\end{eqnarray}
where $\GG^{(h-1)}(\kk')^{-1}$ is defined as in eq.(\ref{11}), with $h-1$
istead of $h$,
$\tilde V^h=\LL \tilde V^h+(1-\LL) V^h$,
$$\LL \tilde V^h=\g^h n_h F_\nu^{h}+
(a_h-z_h) F_\a^{h}
+
g_{2,h} F^{h}_2+g_{4,h} F^{h}_4$$
and
$$Z_{h-1}(k')=Z_{h}+C_h^{-1} Z_h z_h$$
$$Z_{h-1}(k')\sigma_{h-1}(k')=Z_h\sigma_h(k') +Z_h C_h^{-1} s_h$$
This means that we extract from the effective potential the terms
leading to a mass and wave function renormalization.
Now one can perform the integration respect to $\psi^{(h)}$,
rescaling the effective potential
$\hat V^h(\psi)=\tilde V^h(\sqrt{Z_h\over Z_{h-1}}\psi)$ and
$$\LL \hat V^h=\g^h \nu_h F_\nu^{h}+
\d_h F_\a^{h}+
\l_{2,h} F^{h}_2+
\l_{4,h} F^{h}_4$$
with $\nu_h={Z_h\over Z_{h-1}}n_h$,
$\d_h={Z_h\over Z_{h-1}} (a_h-z_h)$,
$\l_{i,h}=({Z_h\over Z_{h-1}})^2 g_{i,h}$ and $\vec
v_k=\nu_h,\d_h,\l_h$.
We can rewrite eq.(18) as
\begin{eqnarray}\label{neo7}
&&\int {\cal D}\psi^{(\leq h-1)}
e^{-\int dk' \vec\psi^{(\leq
h)+}_{k'}C_h Z_{h-1}(k')\GG^{(h-1)}(k')^{-1}
\vec\psi^{(\leq h)-}_{k'}}\nn\\
&& \int {\cal D}\psi^{(h)}
e^{-\int dk' \vec\psi^{(
h)+}_{k'}\tilde f^{-1}_h Z_{h-1}(k')\GG^{(h-1)}(k')^{-1}
\vec\psi^{(h)-}_{k'}}\nn\\
&&e^{-\hat V^h(\sqrt{Z_{h-1}} \psi^{(\leq h)})}]\nn
\end{eqnarray}
and the
integration of $\psi^{(h)}$ has propagator
$$g^h_{\oo,\oo'}(x-y)={1\over Z_{h-1}}
\int d\kk' e^{i\kk'(x-y)}\tilde f_h(\kk')
\GG^{(h-1)}(\kk')_{\oo,\oo'}$$
with $\GG^{(h-1)}(\kk')$ given by
\begin{displaymath}
{1\over A_h}\left(
\begin{array}{c c}
(-ik_0+\o k')& -\s_{h-1}(k')\\
-\s_{h-1}(k')&(-ik_0-\o k') )
\end{array}\right)
\end{displaymath}
$$A_h=-k_0^2-\kk'^2 -\s_{h-1}(k')^2$$
and $Z_{h-1}\equiv Z_{h-1}(0)$ and 
$$\tilde f_h=Z_{h-1}[{C_h^{-1}\over
Z_{h-1}(k')}-{C_{h-1}^{-1}\over Z_{h-1}}]$$
The result of this integration is in the same form as eq.(\ref{11})
with $h$ replaced by $h-1$,
and we can iterate.

Let we explain the main motivations of the integration
procedure discussed above. In a renormalization group
approach one has to identify the relevant, marginal and irrelevant
interactions. By a power counting argument
one sees
that the terms bilinear in the fields are relevant
and the quartic terms (or the bilinear
ones with a derivative respect to
some coordinate acting on the fields) are marginal.
However there are too many kinds of marginal terms, 
depending on the labels $\o_i$ and $\e_i$
on each fields, so that
their renormalization group flow
seems impossible to study. However (see [2], [3] for a similar
procedure)
the power counting can be improved and many
marginal terms are indeed irrelevant; in particular
all the marginal terms with four or two fields
with $\sum_{i}\e_i\not=0$ are indeed irrelevant. The reason is
that such
terms are generated contracting at least 
a non diagonal propagator and such propagators
are smaller than the diagonal ones by a factor $\s_h\g^{-h}$, see eq.(\ref{fon})
below; this 
will be sufficient for improving the power counting, (see the last
paper in [3]
for the proof of a similar statement in the XYZ chain).
Moreover also the marginal terms with $\sum_{i}\e_i\o_ip_F\not=0$
are irrelevant, by momentum conservation considerations.
In fact if $\sum_{i}\e_i\o_i p_F\not=0$
then the momenta of the fermions cannot be all close to the Fermi 
surface; mathematically this means that, for the compact support 
properties of the propagators, there is an $\bar h$ such that
all scattering process involving fermions such that 
$\sum_{i}\e_i\o_i p_F\not=0$
with scale $h\le \bar h$ are
vanishing. 

The {\it relevant terms} are of two kinds;
the $\nu$ terms, reflecting
the renormalization of the Fermi momentum, and the $\s$ terms, related
to the presence of a gap in the spectrum. 
The presence of the $\nu$ terms is due to the renormalization of the
chemical potential, and in general one
introduces a {\it counterterm}
in the hamiltonian to fix the Fermi momentum, see [2].
In this case however there is no necessity
of adding this counterterm; roughly speaking, $\mu$ can vary in the gap
whithout changing the Fermi momentum {\it i.e.} the position of the
singularity of the propagator. This is a {\it crucial} 
point: {\it if 
one had to put a
$\D$-dependent counterterm in the hamiltonian, then considering
${\partial E_0(\D)\over\partial\D}$ one would be forced to derive also
such
counterterm, and a much more complex BCS equation would appear}.
Regarding the other relevant term, they are related to the BCS gap
generation. However {\it due to the interaction the BCS gap has a non
trivial flow}, so that one has to perform 
different Bogolubov transformations at each integration.

Regarding the
marginal terms, there is an anomalous wave function
renormalization which one has to take into account, what is expected as if
$\D=0$ the theory is a Luttinger liquid.
In general the flow of the marginal terms can be controlled
using some cancellations due to the fact that the Beta function
is "close" (for small $u$) to the Mattis model Beta function. 
In eq.(\ref{lut})
we write the propagator as the Mattis model propagator
plus a remainder, so that the Beta function is equal to the Mattis
model Beta function plus a "remainder" which is small if $\s_h\g^{-h}$
is small.

Let be
$h^*={\rm inf_h}\{\g^{h}\geq |\s_h|\}$.
Note that, if $h^*$ is finite uniformly in $L,\b$ so that
$|\s_{h^*-1}|\g^{-h^*+1}\ge 1$ one has
$$|g^{< h^*}(\xx)|\le {1\over Z_{h^*}}
{C_M\g^{h^*}\over 1+(\g^{h^*}|\xx|)^M}$$
Moreover if $h\ge h^*$ we have
$$|g^h_{\o,\o}(\xx)|\le
{1\over Z_{h}}
{C_M\g^{h}\over 1+(\g^{h}|\xx|)^M}$$
and
\begin{equation}\label{fon}
|g^h_{\o,-\o}(\xx)|\le
{1\over Z_{h}}{|\s_h|\over \g^h}
{C_M\g^{h}\over 1+(\g^{h}|\xx|)^M}
\end{equation}

Moreover for $h\ge h^*$ the bound for the non diagonal
propagator
has a factor more ${|\s_h|\over \g^h}$ with respect to the diagonal
propagator. This is the reason for which the quartic terms with $\sum_i\e_i\not=0$
are irrelevant, despite dimensionally marginal.
Finally
\begin{equation}
g^h_{\oo,\oo}(x-y)=g^h_{\oo,L}(x-y)+C^h_{2,\oo}(x-y)\label{lut}
\end{equation}
with
$$g^h_{\oo,L}(x-y)=\int d\kk' {e^{i\kk'x}\over {Z_h}}
\frac{f_h(\kk')}{-ik_0+
\oo k'}$$
which is just the propagator ``at scale h'' of
the Mattis model, and the other term verify the bound of
$g^h_{\oo,\oo}(\xx;\yy)$ with an extra factor 
${|\sigma_h|\over \g^h}$.

We see from the above bounds that the propagator of the
integration of
all the scale between $h^*$ and $h_\b$\footnote{
Of course if $h^*\le h_\b$ there is no such integration.} 
has the same bound as the
propagator
of the integration of a single scale greater than $h^*$; this will be
used to perform the integration of all the scales $< h^*$ in a single
step, {\it i.e.} integrating directly $\psi^{(< h^*)}$.
In fact $\g^{h^*}$ is a momentum scale and, roughly speaking,
for momenta bigger than $\g^{h^*}$ the theory is "essentially"
a massless theory (up to $O(\s_h\g^{-h})$ terms) while for momenta
smaller than $\g^{h^*}$ is a "massive" theory with mass $O(\g^{h^*})$.

One can prove that the effective potentials $V^h$ 
are well defined, if the running coupling constants are small enough.
More precisely, let we write eq.(14) in coordinate space and let be
$\tilde W^k_n$ the corresponding kernel; it holds that

{\bf Lemma}:{\it Assume that $h^*$ is
is finite uniformly in $L,\b$ and
that for any $h>k\ge h^*$
there exists an $\e$ such that $|\vec v_h|\le\e$
and $|{\s_{h+1}\over\s_{h}}|\le \g^{c_a\e}$,
$|{Z_{h+1}\over Z_{h}}|\le \g^{c_b\e^2}$ with $c_a,c_b$
postive constants.
Then there exist a constant $C$ such that
$$||\tilde W_n^k||\le N\b\e C\g^{-k({n\over 2}-2)}$$}

The proof of the above lemma is an immediate modification of ones
existing in literature, see in particular [3].

In order to prove that the effective potentials are well defined
we have to show that the above conditions of smallness in the above lemma
on the running coupling constants are verified.

The beta function for $\nu_h$ is
$$\nu_{h-1}=\g\nu_h+\b_h+\tilde\b_h$$
where $\b_h$ is the contribution
to $\nu_h$ obtained setting
$\s_{k'}=0$, $0\le k'\le h$, so it is exactly equal to $0$
by the parity properties of the Mattis model,
and
$$|\tilde\b_h|\le C{\s_h\over\g^h}\l^2$$
is the remaining part.
Iterating the above relation we find
$$\nu_{h-1}=\g^{|h|}\sum_{k=0}^h \g^k\tilde\b_k\le
\g^{|h^*|}\l^2\sum_k |\s_k|\le C|\l|$$
as $\s_k\simeq \D \g^{\h_1 k}$, $\h_1=-\b_1\l+O(\l^2)$, see below.

The Beta function can be written, for $0\geq h\geq
h^*$. :
\begin{eqnarray}\label{A2}
&&\l_{2,h-1}=\l_h+G^{1,h}_2+G^{2,h}_{2}\nn\\
&&\l_{4,h-1}=\l_h+G^{1,h}_4+G^{2,h}_{4}\nn\\
&&\sigma_{h-1}=\sigma_h
+G_{\sigma}^{1,h}\nn\\
&&\d_{h-1}=\d_h+G^{1,h}_\d
+G^{2,h}_{\d}\nn\\
&&\frac{Z_{h-1}}{Z_h}=1+G_{z}^{1,h}+
G^{2,h}_{z}
\end{eqnarray}
where ($i=2,4$)

a) $G^{1,h}_i$,
$G^{1,h}_\d$
and
$G^{1,h}_z$ depend only on $\l_{i,0},\d_0;...\l_{i,h},\d_h$ and
are given by series
of terms involving only the Mattis model part of the propagator
$g^k_{\oo,L}(x-y)$, so they coincide with the Mattis model Beta 
function 

b)
$G^{2,h}_\sigma$, $G^{2,h}_i,G^{2,h}_\d, G^{2,h}_z$
are given by a series
of terms involving at least a propagator $C_{2,\oo}^k(x-y)$ or
$g^{k}_{\oo,-\oo}(x-y)$ with $k\geq h$.

By a simple explicit computation
$$G^{1,h}_z=\l_h^2[\b_2+\bar G_z^h]$$
$$G^{1,h}_\sigma=\l_h\s_h[\b_1 +\bar G^h_\s]$$
with $\b_1,\b_2>0$ and $\bar G_z^h,\bar G^h_\s=O(\l_h)$.
{\it Moreover $G^{1,h}_i$, $G^{1,h}_\d$ coincide by definition with the
Mattis model Beta function, and it was proved in [2],[3]
that it is vanishing at any order}, {\it i.e.}
$$G^{1,h}_i=
G^{1,h}_\d=0$$
Finally as
$|G_i^{2,h}|,|G_\d^{2,h}|,|G_z^{2,h}|\leq K\e^2 |\s_h|\g^{-h}$,
one finds, for $h\ge h^*$,
$$|\l_{i,h-1}-\l_{i,0}|<c_1 \l^2\quad|\d_{h-1}-\d_0|\le c_1 \l^2$$
and $\s_h\simeq (\D)\g^{\h_1 h}$, $Z_h\simeq \g^{-\h_2 h}$ for
$h\ge h^*$, with $\h_1=-\b_1\l+O(\l^2)$, $\h_2=\b_2\l^2+O(\l^3)$.
As usual in models to which the RG is succesfully
applied the flow is essentially {\it described
by the second order truncation of the
beta function}. This shows that it is possible to choose $\l$ so small
that the conditions of the above lemma are fulfilled. From the
definition of $h^*$ and the fact that 
$\s_h\simeq (\D)\g^{\h_1 h}$ it follows
$\s_{h^*}=\D^{1-\b_1\l+O(\l^2)}$.


As we said
the integrations of the
$\psi^{(< h^*)}$ (if $h^*\ge h_\b$)
is essentially {\it equivalent to the integration
of a single scale $h\ge h^*$}, so it is well defined
by the preceding
arguments. If $h^*< h_\b$ there is no such integration,
and the last scale to be integrated is 
$h_\b$; from this consideration
one obtains easily that the critical temperature is proportional
as usual to the gap amplitude.

An expansion for the two points Schwinger function can be derived in a
standard way [2] and
from the proof of the convergence of the expansion for the effective
potential one obtains easily the convergence of the series for the
Schwinger function. We can write
the r.h.s. of eq.(4) as
\begin{equation}\label{ma}
\sum_{h=\max[h^*,h_\b]}^0 {(2\pi)^2\over L\b}\sum_{\kk'} {\s_h\over
Z_h}{f_h(\kk')\over k_0^2+k'^2+\s_h^2}[1+\l\bar S_h(\kk')]
\end{equation}
where $S_h(\kk')$ is a convergent series bounded by a constant; from the
above expression one can easily derive the BCS equation eq.(5)
as well as the critical temperature.

Note the crucial role of the renormalization of the BCS gap $\s_h$; it
is sensitive to the sign of the interaction and it eliminates or 
enlarges
the singularity of the r.h.s. of eq.(5); neglecting such renormalization
one obtains completely
different results. 
In fact our model belongs to the class of universality
of the {\it massive Luttinger model}, for which it is well known that
the bare mass $\D$ is renormalized 
by the interaction to be given by
$\D^{1-\h}$, $\h=O(\l)$; other models
belonging to this class are the $XYZ$ chain or the Yukawa$_2$ model, see
[3],[9].
Note also that no role is plaid in the above analysis by the {\it spin-charge}
separation; in fact $|v_c-v_s|=O(\l)$ and such effect is incorporated in
the term $\bar S^h$ in eq.(\ref{ma})

\section{\bf References}

\begin{description}
\item[[1]] P.W. Anderson {\it The theory of high $T_c$ superconductivity},
Princeton Press, 1997; and reprints by various authors therein 

\item[[2]] G. Benfatto, G. Gallavotti, A. Procacci,
                  B. Scoppola Comm. Math. Phys {\bf 160}, 93-171;

\item[[3]]F. Bonetto, V. Mastropietro, 
Comm. Math. Phys. {\bf 172}, 57-93 (1995); 
Phys. Rev. B 56 1296-1308 (1997);
Nucl. Phys. B 497 541-554
(1997); V.Mastropietro, 
in print on Comm .Math .Phys.; V.Mastropietro,
submitted to Math. Phys. Lett.

\item[[4]] D. Mattis, {\it Physics}, 1, 183-193 (1964)

\item[[5]] A.M. Tsvelik, Quantum field theory
in condensed matter physics,
Cambridge Un. Press, 1997

\item[[6]] G.Benfatto, G. Gentile, V.Mastropietro, Jour, Stat. Phys.
92,314,
1998

\item[[7]] K.Penc, J.Solyom: Phys. Rev. B, {\bf 41},1, 704-716 (1990);
M.Fabrizio, A. Parola, E. Tosatti, {\bf 46}, 5, 3159-3162 (1992);
A.A. Nerseyan, A. Luther, F.V.Kusmartsev, Physics Letters A, 176, 363-370
(1993)

\item[[8]] Solyom, Advances in Physics, 28, 201-303 (1978)

\item[[9]] A.Luther, Phys. Rev. B, 14,5, 2153-2159 (1976)

\end{description}
\end{document}